\documentstyle[pre,aps]{revtex}

\begin{document}

\draft

\title{Renormalization--Group Solutions for Yukawa Potential}

\author{V.I. Yukalov$^{1,2}$, E.P. Yukalova$^{1,2}$, and F.A. Oliveira$^2$}

\address{$^1$Centre for Interdisciplinary Studies in Chemical Physics \\
University of Western Ontario, London, Ontario N6A 3K7, Canada \\
$^2$International Centre of Condensed Matter Physics\\
University of Brasilia, CP 04513, Brasilia, DF 70919--970, Brazil}

\maketitle

\begin{abstract}

The self--similar renormalization group is used to obtain expressions for 
the spectrum of the Hamiltonian with the Yukawa potential. The critical 
screening parameter above which there are no bound states is also 
obtained by this method. The approach presented illustrates that one can 
achieve good accuracy without involving extensive numerical calculations, 
but invoking instead the renormalization--group techniques.

\end{abstract}

\vspace{2cm}

\pacs{03.65. Ge, 03.65. Db}

\section{Introduction}

Renormalization--group techniques are widely used in quantum field 
theory, statistical mechanics, and solid--state physics. Their usage in 
atomic physics is less customary. The aim of the present paper is to show 
how renormalization--group ideas can be applied for calculating the 
spectra of quantum--mechanical Hamiltonians with realistic potentials.

As a model for illustration we opt for a Hamiltonian with the Yukawa 
potential. This choice is based on the special role of this potential in 
different branches of physics. In plasma physics it is known as the 
Debye--H\"uckel potential, in solid--state physics and atomic physics it 
is called the Thomas--Fermi or screened Coulomb potential, and in nuclear
physics one terms it the Yukawa potential. Among recent important 
applications of this potential we may mention its use in the models 
describing metal--insulator transition [1].

The problem of finding the energy levels for the Yukawa potential has 
received a lot of attention in literature. Several approaches have been 
used for solving this problem: the simple variational procedure [2], use 
of atomic orbitals with a set of fitting parameters [3], 
Rayleigh--Schr\"odinger perturbation theory [4,5], method of potential 
envelopes [6-9], an iterative procedure [10], and different numerical 
calculations [11-14].

In this paper we demonstrate how the problem can be treated by employing 
the self--similar approximation theory [15-19] based on the ideas of the 
renormalization--group and dynamical theory. The outline of the paper is 
as follows. In Sec.II we sketch the main steps of the procedure using 
the self--similar renormalization group. We present only those formulas 
that are necessary for understanding the following calculations; all 
details and mathematical foundation can be found in Refs.[15-19]. In 
Sec.III we apply the approach to the Schr\"odinger equation with the 
Yukawa potential. In Sec.IV we obtain the sequence of renormalized 
energies. The convergence of this sequence is governed by control 
functions defined from the minimum of multiplier. In Sec.V the procedure 
is applied to calculating the renormalized critical screening parameter. 
Emphasize that we are presenting here an analytical method, not relying 
on heavy numerical calculations. Despite its analytical nature, the 
method gives quite good accuracy for the found critical screening 
parameter.

\section{Self--Similar Renormalization} 

We give here the general sketch of the procedure [15-19], without 
specifying the nature of the functions involved. Suppose a function 
$f(x)$ is defined by a complicated equation that is being solved 
approximately. Employing a perturbative algorithm, we may get a sequence 
of approximations, $\{ F_k\}_{k=0}^\infty$, for the sought function 
$f(x)$. To make the sequence convergent, we incorporate into the 
approximations $F_k=F_k(x,u_k)$, with $k=0,1,2,\ldots$, a set of control 
functions $u_k=u_k(x)$, so that the sequence $\{ f_k(x)\}_{k=0}^\infty$ 
of the approximations
\begin{equation}
f_k(x)\equiv F_k(x,u_k(x))
\end{equation}
be convergent.

Make the change of variables by defining a function $x_k(f)$ through the 
equation
\begin{equation}
F_0(x,u_k(x))=f, \qquad x=x_k(f),
\end{equation}
in which $f$ is the new variable. With this change of variables, Eq.(1) 
yields
\begin{equation}
y_k(f)\equiv f_k(x_k(f)) .
\end{equation}
The transformation inverse to Eq.(3) is
\begin{equation}
f_k(x)=y_k(F_0(x,u_k(x)) .
\end{equation}

Construct an approximation cascade $\{ y_k\}$ by requiring the 
self--similarity relation
\begin{equation}
y_{k+p}(f)=y_k(y_p(f)) .
\end{equation}
The trajectory $\{ y_k(f)\}_{k=0}^\infty$ of this approximation cascade 
is, according to Eqs.(3) and (4), bijective to the sequence $\{ 
f_k(x)\}_{k=0}^\infty$ of approximations in Eq.(1). Embedding the 
approximation cascade into an approximation flow and integrating the 
evolution equation for the latter, we obtain the evolution integral
\begin{equation}
\int_{f_k}^{f_{k+1}^*}\frac{df}{v_k(f)} =t_k^* ,
\end{equation}
in which $f_k=f_k(x)$ is a given approximation, $f_k^*=f_k^*(x)$ is a 
renormalized self--similar approximation, and
\begin{equation}
v_k(f) =F_{k+1}(x_k,u_k) -F_k(x_k,u_k) +(u_{k+1}-u_k)
\frac{\partial}{\partial u_k} F_k(x_k,u_k)
\end{equation}
is the cascade velocity, where $x_k=x_k(f)$ and $u_k=u_k(x_k(f))$. The 
right--hand side of Eq.(6), that is $t_k^*$, is the minimal time 
necessary for reaching the renormalized approximation $f_{k+1}^*(x)$.

A fixed point $y_k^*(f)$ of the approximation cascade represents, by 
construction, the sought function $f(x)$ which can be obtained from 
transformation (4). At the fixed point the cascade velocity 
$v_k(f)\rightarrow 0$ as $k\rightarrow\infty$. If the cascade velocity is 
not zero exactly, but only approximately, then we have not an actual fixed
point, but a quasifixed point. For instance, assuming that $v_k(f)\approx 0$
and $F_{k+1}\approx F_k$, from Eq.(7) we have
\begin{equation}
(u_{k+1} -u_k)\frac{\partial}{\partial u_k} F_k(x,u_k) = 0 .
\end{equation}
which is a quasifixed--point condition.

The convergence of the approximation sequence $\{ f_k(x)\}_{k=0}^\infty$ 
is equivalent to the stability of the cascade trajectory $\{ 
y_k(f)\}_{k=0}^\infty$. The stability of the latter can be analysed by 
defining the multipliers
\begin{equation}
\mu_k(f)\equiv\frac{\partial}{\partial f} y_k(f)
\end{equation}
and
\begin{equation}
m_k(x) \equiv\frac{\delta F_k(x,u_k(x))}{\delta F_0(x,u_k(x))} .
\end{equation}
These multipliers are images of each other being related by the 
transformations
\begin{equation}
\mu_k(f) =m_k(x_k(f)) , \qquad m_k(x) =\mu_k(F_0(x,u_k(x))) .
\end{equation}
The trajectory is locally stable when
\begin{equation}
|m_k(x)|\leq 1, \qquad |\mu_k(f)| \leq 1 .
\end{equation}
The multipliers (9) and (10) describe the local stability, at the step 
$k$, of the cascade trajectory with respect to the variation of initial 
conditions. This type of local multipliers can be called quasilocal 
multipliers [19]. Another type of local multipliers defined as
$$ m_k^*(x) \equiv\frac{m_k(x)}{m_{k-1}(x)} $$
characterizes the local stability, at the step $k$, with respect to the 
variation of the point $k-1$. The latter multipliers can be termed 
ultra--local.

Recall that control functions are introduced so that to provide the 
convergence of the approximation sequence $\{ f_k(x)\}_{k=0}^\infty$, 
that is, the stability of the cascade trajectory $\{ y_k(f)\}_{k=0}^\infty$.
This suggests a way for the practical definition of control functions. The 
local multiplier (10) may be written as
\begin{equation}
m_k(x) = M_k(x,u_k(x))
\end{equation}
with
\begin{equation}
M_k(x,u) =\frac{\partial F_k(x,u)}{\partial u} {\Big /}
\frac{\partial F_0(x,u)}{\partial u} . 
\end{equation}
To produce the maximal stability of the cascade trajectory, i.e., the 
fastest convergence of the approximation sequence, for each fixed value of $x$,
we have to require the minimum of the absolute value for the multiplier (13)
with respect to the control function $u_k(x)$. In other words, the 
principal of maximal stability is the condition for the minimum of the 
multiplier modulus,
\begin{equation}
\min_{u}|M_k(x,u)| =|M_k(x,u_k(x))| .
\end{equation}
This condition gives us a constructive definition of control functions.

\section{Yukawa Potential}

Now we illustrate how the scheme of Sec.II applies for calculating the
eigenvalues of the radial Hamiltonian
$$ H = -\frac{1}{2m}\frac{d^2}{dr^2} +\frac{l(l+1)}{2mr^2} -
\frac{A}{r} e^{-\alpha r} $$
with the Yukawa potential.

It is convenient to pass to dimensionless quantities by scaling the above
Hamiltonian and reducing it to the form
\begin{equation}
H =-\frac{1}{2}\frac{d^2}{dr^2} +\frac{l(l+1)}{2r^2} -\frac{e^{-\alpha r}}{r} .
\end{equation}
Here $r\in[0,\infty);\; l=0,1,2,\ldots$; and $\alpha$ is a positive 
constant called the screening parameter. To return back to dimensional 
quantities one has to make the following substitutions:
$$ r\rightarrow mAr , \qquad \alpha\rightarrow\frac{\alpha}{mA}, \qquad
H\rightarrow \frac{H}{mA^2} . $$

Write Hamiltonian (16) as the sum $H\equiv H_0+\Delta H$, with the first 
term being the Hamiltonian
\begin{equation}
H_0 =-\frac{1}{2}\frac{d^2}{dr^2} +\frac{l(l+1)}{2r^2} -\frac{u}{r}
\end{equation}
with a Coulomb--type potential, where $u$ is yet unknown quantity which 
will latter generate control functions.

Employing some variant of perturbation theory in powers of the perturbation
\begin{equation}
\Delta H\equiv H -H_0 = \frac{u-e^{-\alpha r}}{r} ,
\end{equation}
we may construct a sequence of approximate eigenvalues, $E_k$, and 
eigenfunctions, $\psi_k$, respectively,
\begin{equation}
E_k\equiv E_{nl}^{(k)}(\alpha,u), \qquad \psi_k\equiv\psi_{nl}^{(k)}(r,u) ,
\end{equation}
where $k=0,1,2,\ldots$ enumerates approximations, while $n=0,1,2,\ldots$ 
and $l=0,1,2,\ldots$ are the quantum numbers labelling the energy levels. 
For the initial approximation one has the spectrum
\begin{equation}
E_0=-\frac{u^2}{2(n+l+1)^2} ,
\end{equation}
and the wave function
\begin{equation}
\psi_0=\left [\frac{n!u}{(n+2l+1)!}\right ]^{1/2}
\frac{1}{n+l+1}\left (\frac{2ur}{n+l+1}\right )^{l+1}
\exp\left (-\frac{ur}{n+l+1}\right ) 
L_n^{2l+1}\left (\frac{2ur}{n+l+1}\right ) ,
\end{equation}
in which
$$ L_n^l(r) 
=\sum_{m=0}^n\frac{\Gamma(n+l+1)(-r)^m}{\Gamma(m+l+1)(n-m)!m!} $$ 
is an associate Laguerre polynomial.

To find the subsequent approximations, we could use the 
Rayleigh--Schr\"odinger perturbation theory. However this would involve 
the following complication. The whole spectrum of the Hamiltonian (16) 
contains, in addition to discrete levels, the continuous part. Therefore, 
we would have to deal, besides the summation over discrete levels, with 
the integration over continuous ones.

To avoid this complication, when we are interested only in discrete levels, 
we may employ the Dalgarno--Lewis perturbation theory [20]. Then, for 
the $k$--approximation one writes
$$ E_k = E_0 +\sum_{p=1}^k\Delta E_p , $$
\begin{equation}
\psi_k =\psi_0 +\sum_{p=1}^k\Delta\psi_p .
\end{equation}
The first correction for the eigenvalues is
\begin{equation}
\Delta E_1 =(\psi_0,\Delta H\psi_0) ,
\end{equation}
and the first correction to the eigenfunction is a solution to the equation
\begin{equation}
(H_0 -E_0)\Delta\psi_1 =(\Delta E_1 -\Delta H)\psi_0 .
\end{equation}
Solving the Dalgarno equation (24), one may calculate the second 
correction to the eigenvalues,
\begin{equation}
\Delta E_2 =(\Delta\psi_1,\Delta H\psi_0) ,
\end{equation}
and so on.

The Dalgarno equation (24) is a nonhomogeneous equation whose solution 
can be written as the sum of the general solution to the corresponding 
homogeneous equation plus a particular solution to the nonhomogeneous 
equation. The solution to the homogeneous equation is, as is evident, 
proportional to $\psi_0$. So we may set
\begin{equation}
\Delta\psi_1 = C\psi_0 +\varphi ,
\end{equation}
with the proportionality constant $C$ defined by the normalization condition 
$(\psi_k,\psi_k)=1$, and the function $\varphi$ being a particular 
solution to the nonhomogeneous equation
\begin{equation}
(H_0 - E_0)\varphi =(\Delta E_1 -\Delta H)\psi_0 .
\end{equation}
From the normalization condition $(\psi_1,\psi_1)=1$, for the function 
$\psi_1=\psi_0+\Delta\psi_1$, omitting the second--order term, one has
\begin{equation}
(\psi_0,\Delta\psi_1) = 0 .
\end{equation}
With Eq.(26), this gives
\begin{equation}
C=-(\psi_0,\varphi) .
\end{equation}

Following the scheme described, with the notation
\begin{equation}
\beta\equiv\frac{\alpha}{2u}(n+l+1) ,
\end{equation}
we obtain the first correction for the eigenvalues of bound states,
\begin{equation}
\Delta E_1 =\frac{u^2I_0 -uI_\beta}{(n+l+1)^2} ,
\end{equation}
where the integral
$$ I_\beta\equiv\frac{n!}{(n+2l+1)!}
\int_0^\infty r^{2l+1}e^{-(1+\beta)r}\left [ L_n^{2l+1}(r)\right ]^2 dr $$
can be expressed as
$$ I_\beta =\frac{(\beta-1)^n}{(\beta+1)^{n+2l+2}}{\cal P}_n^{2l,0}\left (
\frac{\beta^2+1}{\beta^2-1}\right ) $$
through the Jacobi polynomials
$$ {\cal P}_n^{k,p}(x) 
=\frac{(-1)^n}{2^nn!}(1-x)^{-k}(1+x)^{-p}\frac{d^n}{dx^n}\left [ 
(1-x)^{n+k}(1+x)^{n+p}\right ] = $$
$$ = \frac{1}{2^n}\sum_{m=0}^nC_{n+k}^mC_{n+p}^{n-m}(x-1)^{n-m}(x+1)^m $$
having the properties
$$ {\cal P}_n^{k,p}(1) =C_{n+k}^n, \qquad {\cal P}_n^{k,p}(-1) 
=(-1)^nC_{n+p}^n, \qquad C_n^m \equiv \frac{n!}{(n-m)!m!} . $$
Another integral in Eq.(31) is
$$ I_0\equiv\lim_{\beta\rightarrow 0}I_\beta = 1 . $$

In this way, the first approximation $E_1=E_0+\Delta E_1$ becomes
\begin{equation}
E_1=\frac{u^2-2uI_\beta}{2(n+l+1)^2} ,
\end{equation}
where
$$ I_\beta =\frac{1}{(1+\beta)^{2n+2l+2}}
\sum_{m=0}^n C_{n+2l+1}^m C_n^{n-m}\beta^{2m} . $$
For the ground--state level, when $n=l=0$ and $\beta=\alpha/2u$, Eq.(32) 
reduces to
\begin{equation}
E_1=-u^2\left (\frac{1}{2}-\sigma\right ) ,
\end{equation}
where the notation
\begin{equation}
\sigma\equiv 1-\frac{4u}{(2u+\alpha)^2}
\end{equation}
is introduced.

The ground state plays a special role defining, when it becomes zero, the
critical screening parameter $\alpha_c$, above which there are no bound 
states. Therefore, in what follows we consider the ground state.

Eq.(27), with
\begin{equation}
E_0=-\frac{u^2}{2}, \qquad \psi_0 =2u^{3/2}re^{-ur} ,
\end{equation}
writes
\begin{equation}
\left (-\frac{1}{2}\frac{d^2}{dr^2} -\frac{u}{r}+\frac{u^2}{2}\right )\varphi
= \left (\sigma u^2 -\frac{u}{r} +\frac{e^{-\alpha r}}{r}\right )\psi_0 .
\end{equation}
The solution to Eq.(36) must be a bounded function, $|\varphi(r)|<\infty$, 
for all $r\in[0,\infty)$.

Let us present $\varphi$ as the product
$$ \varphi(r) =\psi_0(r)g(ur) , $$
in which the second factor satisfies the equation
$$ \frac{d}{dr} g(r) =\rho(r)\frac{e^{2r}}{r^2} , $$
where $\rho(r)$ is to be defined from Eq.(36), which yields
$$ \frac{d\rho}{dr} =(1-\sigma r)2re^{-2r} -
\frac{2r}{u}\exp\left (-\frac{2u+\alpha}{u}r\right ) . $$
The latter equation gives
$$ \rho(r) =\left [ \sigma r^2 +(\sigma -1)\left ( r+\frac{1}{2}\right 
)\right ] e^{-2r} +\frac{1-\sigma}{2}\left ( 1 +\frac{2u+\alpha}{u}r\right )
\exp\left (-\frac{2u+\alpha}{u}r\right ) + C_1, $$
with an integration constant $C_1$. The equation for $g(r)$ results in 
$$ g(ur)=\sigma ur +(1-\sigma)\left [\frac{1-e^{-\alpha r}}{2ur} -\ln(ur) +
Ei(-\alpha r)\right ] +C_1\left [ 2Ei(2ur) -\frac{e^{2ur}}{ur}\right ] +C_2, $$
where $C_2$ is an integration constant and
$$ Ei(ar) =\int_{-\infty}^r\frac{e^{ax}}{x}dx $$ 
is the exponential--integral function.

The boundness of $\varphi$ requires that $C_1=0$. The additive term 
containing a function proportional to $\psi_0$ should be omitted in 
$\varphi$, since such a term has already been included into Eq.(26). The 
latter means that we have to put $C_2=0$. As a result we obtain
\begin{equation}
\varphi(r) =\sqrt{u}e^{-ur}\left\{ 2\sigma(ur)^2 +(1-\sigma)\left [
1 - e^{-\alpha r} - 2ur\ln(ur) +2urEi(-\alpha r)\right ]\right\} .
\end{equation}
For the normalization  constant in Eq.(29) we find
\begin{equation}
C=-\frac{3}{2}\sigma +(1-\sigma)\left (\frac{\alpha}{2u+\alpha} 
-\ln\frac{2\alpha}{2u+\alpha}-\gamma_E\right )
\end{equation}
with the Euler constant $\gamma_E=0.577215$.

After finding function (26), we can calculate the second correction (25),
which yields
\begin{equation}
\Delta E_2=u^2J_1 -u J_2 ,
\end{equation}
where 
$$ J_1=\lim_{\nu\rightarrow 0}J_2(\nu), \qquad J_2=J_2(\beta) , $$
with $\beta=\alpha/2u$, and
$$ J_2(\nu) =2\int_0^\infty\left\{ 2Cr + 2\sigma r^2 +(1-\sigma)\left [ 1 -
e^{-\alpha r/u} - 2r\ln r + 2rEi\left (-\frac{\alpha r}{u}\right )
\right ]\right \} e^{-2(1+\nu)r} dr . $$
Thus, for the second approximation for the energy we obtain
\begin{equation}
E_2 =E_1 +u^2 J_1 -u J_2 ,
\end{equation}
where
$$ J_1 =-\frac{\sigma}{2} +\frac{1-\sigma}{1+\beta}\beta , $$

$$ J_2 = -u(1-\sigma)\frac{\sigma(1+3\beta)}{2(1+\beta)} + 
u(1-\sigma)^2\left [ \frac{(1+3\beta+\beta^2)\beta}{(1+\beta)(1+2\beta)} +
\ln\frac{(1+\beta)^2}{1+2\beta}\right ] . $$

\section{Renormalized Energy}

From the results of the previous section we derive  the following 
sequence of approximations for the energy:
$$ E_0 =-\frac{u^2}{2} , $$
$$ E_1= E_0 +u^2 -\frac{4u^3}{(2u+\alpha)^2} , $$
$$ E_2 =E_1 -\frac{u^2}{2} +\frac{2u^3}{(2u+\alpha)^2} 
+\frac{4u^3\alpha}{(2u+\alpha)^3} 
+\frac{2u^3(2u^2+5u\alpha-2\alpha+3\alpha^2)}{(u+\alpha)(2u+\alpha)^3} - $$
\begin{equation}
-\frac{8u^4(2u^2+5u\alpha+4\alpha^2)}{(u+\alpha)(2u+\alpha)^5} -
\frac{16u^4}{(2u+\alpha)^4}\ln\frac{(2u+\alpha)^2}{4u(u+\alpha)} . 
\end{equation}

For multiplier defined in Eq.(14) we have
\begin{equation}
M_k(\alpha,u) =\frac{\partial E_k(\alpha,u)}{\partial u}{\Big /}
\frac{\partial E_0(\alpha,u)}{\partial u} .
\end{equation}
Substituting here the derivatives
$$ \frac{\partial E_0}{\partial u} = - u, $$
$$ \frac{\partial E_1}{\partial u} = u -\frac{12u^2}{(2u+\alpha)^2} 
+\frac{16u^3}{(2u+\alpha)^3} , $$
following from the sequence (41), we find
\begin{equation}
M_1(\alpha,u) = -
\frac{8u^3-4(2-3\alpha)u^2-6\alpha(2-\alpha)u+\alpha^3}{(2u+\alpha)^3} .
\end{equation}
The control function $u(\alpha)=u_1(\alpha)$ is to be defined from the 
principle of maximal stability (15). To this end, we, first, try the equation
\begin{equation}
M_1(\alpha,u) = 0,
\end{equation}
which gives
\begin{equation}
8u^3-4(2-3\alpha)u^2-6\alpha(2-\alpha)u+\alpha^3 = 0 .
\end{equation}
This cubic equation has three roots of which we have to select a real 
one satisfying the asymptotic condition
$$ \lim_{\alpha\rightarrow 0}u(\alpha) = 1 . $$
The latter implies that if the screening parameter tends to zero, so that 
the Yukawa potential transforms into the Coulomb one, then one must 
return to the exact solution known for the Coulomb potential. Really, 
under $\alpha\rightarrow 0$ and $u\rightarrow 1$, from the sequence in 
Eq.(41) it follows that $\;E_k\rightarrow -\frac{1}{2}$ for all $k$. 
Eq.(45), with this asymptotic condition, yields the control function
\begin{equation}
u(\alpha) =\frac{1}{3} -\frac{\alpha}{2} 
+\frac{2}{3}\sqrt{1+\frac{3}{2}\alpha}\;\cos\frac{\varphi}{3} ,
\end{equation}
in which
\begin{eqnarray}
\varphi=\left\{ \begin{array}{cc}
\varphi^*, & \varphi^*\geq 0, \; 0\leq\alpha\leq\frac{3+\sqrt{57}}{18} , \\
\nonumber
\pi-\varphi^*, & \varphi^*\leq 0, \; 
\frac{3+\sqrt{57}}{18} \leq \alpha \leq \alpha_0 , \end{array}\right.
\end{eqnarray}
where
$$ \varphi^* =
\arctan\frac{3\alpha\sqrt{3(3+20\alpha-27\alpha^2)}}{4+9\alpha-27\alpha^2} , $$
and the upper value of the screening parameter, below which Eq.(45) 
possesses yet
a solution, is
\begin{equation}
\alpha_0\equiv\frac{10+7\sqrt{7}}{27} = 1.056306.
\end{equation}
Thus, solution (46) exists only in the interval 
$0\leq\alpha\leq\alpha_0$. For $\alpha>\alpha_0$, Eq.(45) has no real 
solutions satisfying the derived asymptotic condition.

For $\alpha>\alpha_0$ we need to find a minimum of the multiplier (43), 
which is not necessarily zero. This can be done by solving the equation
\begin{equation}
\frac{\partial}{\partial u}M_1(\alpha,u) = 0 ,
\end{equation}
which results in the control function
\begin{equation}
u(\alpha) =\left ( \frac{\sqrt{7}}{2} -1\right )\alpha = 0.322876\alpha .
\end{equation}

One may notice that when Eq.(44) has a solution, then the procedure is 
similar to renormalizing perturbative terms by means of a variational 
optimization [21-26]. However, as is shown above, this equation not 
always possesses physically reasonable solutions. While the principle of 
maximal stability (15) always provides us with a solution defining a 
control function. Therefore, this principle is more general than the 
simple variational procedure.

Substituting the found control function into the approximations in the 
sequence (41), we get the renormalized expressions
\begin{equation}
e_k(\alpha)\equiv E_k(\alpha,u_k(\alpha)) .
\end{equation}
For example, when $\alpha\leq\alpha_0$, we have
\begin{equation}
e_1(\alpha) =-\frac{u^2(2u-\alpha)}{2(2u+3\alpha)} , 
\end{equation}
\begin{equation}
e_2(\alpha) =-\frac{u^2}{2}\left [
\frac{8u^4+16u^3\alpha-2u^2\alpha^2-10u\alpha^3+\alpha^4}
{2u(u+\alpha)(2u+3\alpha)^2} + 
2 \left (\frac{2u+\alpha}{2u+3\alpha}\right )^2
\ln\frac{(2u+\alpha)^2}{4u(u+\alpha)}\right ] .
\end{equation}
Respectively, for $\alpha>\alpha_0$, we have to substitute the control 
function (49) into Eq.(41).

The self--similar approximation for the energy is to be defined from the 
evolution integral (6). When no additional constraints are imposed, the 
minimal number of steps for reaching a quasifixed point is, clearly, one, 
$t_k^*=1$. In the interval $0\leq\alpha\leq\alpha_0$, the cascade 
velocity, given by Eq.(7), is
\begin{equation}
v_1(f)=e_2(\alpha(f))-e_1(\alpha(f)) ,
\end{equation}
where the function $\alpha(f)$, according to Eq.(2), is defined by the 
equation
\begin{equation}
E_0(\alpha,u(\alpha)) = -\frac{1}{2}u^2(\alpha) = f
\end{equation}
resulting in $\alpha=\alpha(f)$. For the evolution integral (6), we have
\begin{equation}
\int_{e_1}^{e_2^*}\frac{df}{v_1(f)} = 1 ,
\end{equation}
where $e_1=e_1(\alpha)$ and $e_2^*=e_2^*(\alpha)$.

We calculated the values of the renormalized energies $e_1(\alpha),\;
e_2(\alpha)$, and $e_2^*(\alpha)$ as functions of the screening parameter
$\alpha$. To characterize the accuracy of these approximations, it is 
convenient to introduce the maximal percentage errors
$$ \varepsilon_k\equiv \max_{\alpha}
\left [ \frac{e_k(\alpha)-e(\alpha)}{|e(\alpha)|}\right ]\times 100\% , $$
where $e(\alpha)$ is an exact value of the energy. Notice that this 
definition of the maximal error has sense only when $e(\alpha)$ is not close 
to zero. In our case this definition works for $0\leq\alpha\leq\alpha_0$. 
For $\alpha>\alpha_0$, when $e(\alpha)\rightarrow 0$, it is possible to 
redefine the maximal error by shifting the definition of the energy by a
constant [27].

The maximal percentage error, defined as is explained above, is $2\%$ for
$e_1(\alpha)$ and for $e_2(\alpha)$ and $e_2^*(\alpha)$ it is $1\%$. The 
multipliers (10), with $\alpha$ instead of $x$, satisfy the stability 
conditions of Sec.II. The stability of the procedure means its convergence 
which is also evident from the improvement of accuracy.

\section{Critical Screening}

An important quantity characterizing the features of the Yukawa potential 
is the critical screening parameter, that is, such a value of the 
screening parameter $\alpha=\alpha_c$ above which there are no bound 
states. This critical parameter is defined by the condition $e(\alpha_c)=0$.
For each approximation $e_k(\alpha)$ for the ground--state energy there 
exists the corresponding critical parameter $\alpha_k$ given by the equation
\begin{equation}
e_k(\alpha_k) = 0 .
\end{equation}
From the approximations $e_k(\alpha)$ obtained in the previous section, 
we find the sequence of approximations for the critical screening parameters:
\begin{equation}
\alpha_1 = 1 , \qquad \alpha_2 =1.0833 .
\end{equation}
To employ the self--similar renormalization for the sequence 
$\{\alpha_k\}$, as is described in Sec.II, we compose a sequence 
$\{\alpha_k(\lambda)\}$ of the partial sums
\begin{equation}
\alpha_k(\lambda)=\sum_{i=1}^k(\alpha_i-\alpha_{i-1})\lambda^{p_i} ,
\end{equation}
in which $k\geq1$ and $\alpha_0(\lambda)\equiv0\;$. As is clear from Eq.(58),
\begin{equation}
\alpha_k=\lim_{\lambda\rightarrow 1}\alpha_k(\lambda) .
\end{equation}
Then the sequence of approximations
$$ \alpha_1(\lambda)=\alpha_1\lambda^{p_1} , $$
$$ \alpha_2(\lambda) =\alpha_1\lambda^{p_1} + 
(\alpha_2 -\alpha_1)\lambda^{p_2} $$
can be renormalized in the way prescribed by Sec.II. First, we define the 
expansion function $\lambda(f)$ by the equation
$$ \alpha_1(\lambda)=\alpha_1\lambda^{p_1} = f, $$
which gives
$$ \lambda(f) =\left (\frac{f}{\alpha_1}\right )^{1/p_1} . $$
Writing the cascade velocity as
$$ v_1(f) =\alpha_2(\lambda(f))-\alpha_1(\lambda(f)) =
(\alpha_2 - \alpha_1)\lambda^{p_2}(f) , $$
we come to the evolution integral
\begin{equation}
\int_{\alpha_1(\lambda)}^{\alpha_2^*(\lambda)}
\frac{df}{(\alpha_2-\alpha_1)\lambda^{p_2}(f)} = t^* .
\end{equation}

Notice that $\alpha_2^*(\lambda)$ in Eq.(60) depends also on the parameters
\begin{equation}
p \equiv p_1 \geq 0 , \qquad q \equiv \frac{p_2}{p_1} - 1 \geq 0 ,
\end{equation}
so that we may write
\begin{equation}
\alpha_2^*(\lambda) =\alpha_2^*(\lambda,p,q) .
\end{equation}
Integrating Eq.(60), we obtain
\begin{equation}
\alpha_2^*(\lambda,p,q) =\left [
\frac{\alpha_1^{1+q}\lambda^{pq}}
{\alpha_1-q(\alpha_2-\alpha_1)\lambda^{pq}t^*}\right ]^{1/q} .
\end{equation}

To define $p$ and $q$, consider the sequence $\{\bar y_k(f)\}$ of the 
terms $\bar y_k(f)\equiv\alpha_k(\lambda(f))$. Thus, we have
$$ \bar y_1(f) = f , $$
$$ \bar y_2(f) = f +
(\alpha_2-\alpha_1)\left (\frac{f}{\alpha_1}\right )^{q+1} . $$
For the latter sequence we can find the multipliers defined as in 
Eq.(14). As is obvious, $M_1=1$ and
\begin{equation}
M_2(\lambda,p,q) = 1 +(\alpha_2-\alpha_1)(1+q)\lambda^{pq} .
\end{equation}

The values of $p$ and $q$ are to be chosen so that to satisfy the 
principle of maximal stability (15), with $p$ and $q$ playing the role of 
control functions. Since, according to condition (59), we must put at the 
end $\lambda\rightarrow 1$, we can consider the multiplier (64) for 
$\lambda\sim 1$. In the case when $\lambda>1$, the minimum of $|M_2|$ from 
Eq.(64) is provided by $q=0$. But if $\lambda<0$, then this minimum can occur
at $q=1$. Here we keep in mind that $q$, given by Eq.(61), is an integer 
and that the difference $\alpha_2-\alpha_1$ is positive in agreement with 
Eq.(57). In this way, from expression (63) we derive
\begin{equation}
\alpha_2^*(\lambda,p,0) =\alpha_1\lambda^p\exp\left (
\frac{\alpha_2-\alpha_1}{\alpha_1}t^*\right ) ,
\end {equation}
if $q=0$, and
\begin{equation}
\alpha_2^*(\lambda,p,1)=
\frac{\alpha_1^2\lambda^p}{\alpha_1-(\alpha_2-\alpha_1)\lambda^pt^*} ,
\end{equation}
when $q=1$.

The effective time $t^*$ has the meaning of the minimal number of steps 
providing the renormalization of $\alpha_k$, when $\lambda\rightarrow 1$. 
If we put $\lambda\rightarrow 1$ in the evolution integral (60) before 
the integration, then for $\alpha_2^*$ we would get 
$\alpha_1+(\alpha_2-\alpha_1)t^*$. From here we see that $t^*=0$ gives 
$\alpha_1$; one step, that is $t^*=1$, leads to $\alpha_2$; and $t^*=2$
results in $2\alpha_2-\alpha_1$. Therefore, the minimal number of steps 
necessary for getting a nontrivial renormalization is $t^*=2$.

Eqs. (65) and (66) show that, when $\lambda=1$, then $\alpha_2^*$ does 
not depend on $p$. Consequently, we may write
\begin{equation}
\alpha_2^*(1,q) \equiv \alpha_2^*(1,p,q) .
\end{equation}
Putting in Eqs.(65) and (66) $t^*=2$ and $\lambda=1$, we obtain
\begin{equation}
\alpha_2^*(1,0) =\alpha_1\exp\left ( 
2\frac{\alpha_2-\alpha_1}{\alpha_1}\right )
\end{equation}
and, respectively,
\begin{equation}
\alpha_2^*(1,1) =\frac{\alpha_1^2}{3\alpha_1-2\alpha_2} .
\end{equation}
As the final answer we set
\begin{equation}
\alpha_2^* =\frac{1}{2}\left [ \alpha_2^*(1,0) +\alpha_2(1,1)\right ] .
\end{equation}

Substituting into formulas (68) and (69) the numerical values from 
Eq.(57), we have
$$ \alpha_2^*(1,0)=1.1813, \qquad \alpha_2^*(1,1)=1.1919 . $$
Thence, Eq.(70) yields
$$ \alpha_2^* = 1.1906 . $$
This value of the critical screening parameter coincides with the result 
of numerical integration [11].

In conclusion, we have applied the self--similar renormalization theory 
[15-19] to calculating the energy and the critical screening parameter 
for the Schr\"odinger equation with the Yukawa potential. The calculated 
values are in good agreement with the results of numerical computation. 
This demonstrates that renormalization--group techniques can be successfully
employed for solving the Schr\"odinger equation, not only with simple 
anharmonic potentials [28,29], but also for more realistic cases, such as 
the Yukawa potential that is often met in different physical problems. To 
get an accurate value of the critical screening parameter, we have used a 
method analogous to the algebraic self--similar renormalization [30]. We 
paid here attention mainly to the ground--state level, although the 
procedure we demonstrated is applicable to excited levels as well, but
calculations become a little more cumbersome. We hope that the results we 
obtained are sufficient for showing the usefulness of renormalization--group
techniques in quantum mechanics.

\vspace{5mm}

{\bf Acknowledgement}

\vspace{2mm}

We acknowledge financial support of the National Science and Technology 
Development Council of Brazil and from the University of Western Ontario, 
Canada..

\end{document}